\documentclass[aip,pop,amsmath,amssymb,reprint,color]{revtex4-1}

\usepackage{graphicx}
\usepackage{color}
\usepackage{dcolumn}
\usepackage{bm}

\draft 

\begin{document}

\title{Dynamic Procedure for Filtered Gyrokinetic Simulations} 

\author{P. Morel}
\email[]{pmorel@ulb.ac.be}
\author{A. Ba\~n\'on Navarro}
\author{M. Albrecht-Marc}
\author{D. Carati}
\affiliation{Statistical and Plasma Physics Laboratory, Universit\'e Libre de Bruxelles, Belgium.}

\author{F. Merz}
\author{T. G\"orler}
\author{F. Jenko}
\affiliation{Max-Planck-Institut f\"ur Plasmaphysik, EURATOM Association, D-85748 Garching, Germany}

\date{\today}

\begin{abstract}

Large Eddy Simulations (LES) of gyrokinetic plasma turbulence are investigated as interesting candidates to decrease the computational cost. A dynamic procedure is implemented in the GENE code, allowing for dynamic optimization of the free parameters of the LES models (setting the amplitudes of dissipative terms). Employing such LES methods, one recovers the free energy and heat flux spectra obtained from highly resolved Direct Numerical Simulations (DNS). Systematic comparisons are performed for different values of the temperature gradient and magnetic shear, parameters which are of prime importance in Ion Temperature Gradient (ITG) driven turbulence. Moreover, the degree of anisotropy of the problem, that can vary with parameters, can be adapted dynamically by the method that shows Gyrokinetic Large Eddy Simulation (GyroLES) to be a serious candidate to reduce numerical cost of gyrokinetic solvers.

\end{abstract}


\maketitle 


\section{Motivation and context}

In the area of fluid turbulence, theories are usually based on the notion of an inertial range in which the energy cascades from larger scales to (somewhat) smaller scales mediated by the quadratic nonlinearity. The role of the smallest scales is then to dissipate energy in the so-called dissipative range. In numerical simulations, this picture has led to the development of Large Eddy Simulation (LES) techniques that are based on the idea that neglecting the small scales can be compensated by introducing a dissipative model for the eddy viscosity~\cite{smagorinsky-MWR-1963}.

A Direct Numerical Simulation (DNS) is supposed to retain all the scales from the injection range down to the dissipative range. This requires an enormous numerical effort in the case of high Reynolds number flows. On the contrary, a LES coarsens the simulation grid and only retains the largest scales (which are problem-dependent), while the small scales (which are assumed to be universal) are replaced by a model. In Fourier space, such a coarsening can be seen as the action of a low-pass filter. Since the scale range is truncated, the dissipation scales can not be reached, and the modeling basically consists of the introduction of artificial dissipation mechanisms. From a more mathematical viewpoint, one notes that the filtering operation does not commute with the nonlinear term that transfers energy from largest to smallest scales, and the major problem of LES consists in finding a satisfying closure for representing the influence of the unresolved scales.


Recent gyrokinetic studies have shown that Ion Temperature Gradient (ITG) driven turbulence exhibits a direct and local cascade of a nonlinear invariant, namely the free energy.\cite{banon-PRL-2011} Such a cascade is analogous to the kinetic energy cascade in three dimensional Navier-Stokes turbulence. The important difference is that the quadratic conserved quantity in fluid dynamics is the kinetic energy, while it is the free energy in gyrokinetics. The latter quantity is the sum of both the perturbed entropy and the electrostatic energy. Transfers between entropy and electrostatic energy are ensured by the magnetic curvature and parallel dynamics terms.\cite{banon-PoP-2011}

Thus, adapting LES methods to gyrokinetics is quite a natural idea. In a pioneering study, Smith and Hammett have applied LES techniques to a set of gyrofluid equations.\cite{smith-PoP-1997} The use of a hyper-viscosity model was found to provide better agreement than a simple Smagorinsky-type eddy viscosity.\cite{smagorinsky-MWR-1963} Promising recent comparisons between gyrokinetic LES and highly resolved DNS have motivated the present work.\cite{morel-PoP-2011} As is well known, LES approaches have to face two distinct difficulties. First, a suitable model has to be designed to mimic the dissipative effect of the small scales. Second, the free parameters of the model have to be determined in order to ensure that it creates the correct amount of dissipation. In a previous study~\cite{morel-PoP-2011}, the first difficulty has been addressed: the feasibility of Gyrokinetic Large Eddy Simulations (GyroLES) has been demonstrated with perpendicular hyper-diffusion models. However, the second difficulty remains to be tackled. In practice, up to now, the free parameter setting the amplitude of the dissipative term modelling the influence of the neglected scales had to be determined via a trial and error process. The main objective of the present study is to overcome this problem by adapting the dynamic procedure~\cite{germano-JFM-1992,germano-PoF-1991} to GyroLES. The dynamic procedure is an optimization approach that allows to calibrate the model amplitude in the course of the LES.

The remainder of the present paper is organized as follows. After a brief review of the GyroLES formalism in Section \ref{sec:gyroLES}, the effect of truncating small scales is studied in detail, and the dynamic procedure for gyrokinetics is discussed in Section \ref{sec:model-design}. Numerical results obtained for various logarithmic temperature gradient and magnetic shear values are presented in Section \ref{sec:numerics}.

\section{LES formalism in gyrokinetics \label{sec:gyroLES}}

In the following, the nonlinear gyrokinetic equations are solved by means of the GENE code \cite{gene}. Although a more comprehensive code version including nonlocal effects is at hand \cite{goerler11}, for simplicity we restrict ourselves here to the local code version. Only electrostatic fluctuations are considered, with a fixed background magnetic field $B_0$ and adiabatic electrons. Field aligned coordinates are used \cite{beer-PoP-1995}, with the assumption of circular concentric flux surfaces \cite{lapillonne-PoP-2009}. The GENE code uses a delta-f splitting of the unknown distribution function: $F_i = F_{0i} + f_{ki}$ with the normalized equilibrium distribution function $F_{0i} = e^{- v_\parallel^2 - \mu B_0}$, where $\mu = m_i v_\perp^2 / (2 B_0)$ is the ion magnetic moment (mass $m_i$), $v_\perp$ and $v_\parallel$ are respectively the velocity coordinates perpendicular and parallel to the magnetic field. Unknowns are Fourier transformed along coordinates perpendicular to the magnetic field $(x,y) \rightarrow (k_x , k_y)$. The gyrokinetic Vlasov equation for ions guiding center distribution function $f_{ki} (k_x,k_y,z,v_\parallel,\mu,t)$ then reads:

\begin{equation} 
\partial_t f_{ki} = L [f_{ki}] + N [\phi_{k}, f_{ki}] - D [f_{ki}],
\label{eq:gyrokinetic}
\end{equation} where $L$ represents linear terms, $N$ the quadratic nonlinearity, and $D$ the numerical dissipation terms.

The linear terms can be written as $L = L_{B_0} + L_G + L_\parallel$, where $L_{G} [f_{ki}]$ is the drive due to logarithmic density and temperature gradients ($\omega_{ni}$ and $\omega_{Ti}$), $L_{B_0} [f_{ki}]$ corresponds to both the curvature and the gradient of the magnetic field $B_0$ (referred to as ``curvature'' in the following), and $L_\parallel [f_{ki}]$ is the term describing the parallel dynamics:

\begin{eqnarray}
L_{G} [f_{ki}] & = & - \left [ \omega_{ni} + \left ( v_\parallel^2 + \mu B_0 - \frac{3}{2} \right ) \omega_{Ti} \right ] F_{0i} i k_y J_{0k} \phi_k \, , \nonumber \\
\label{def:LG}
& & \\
\label{def:LB}
L_{B_0} [f_{ki}] & = & - \frac{T_{i0} ( 2 v_\parallel^2 + \mu B_0 )}{Z_i T_{e0} B_0} \left [ K_x i k_x + K_y i k_y \right ] h_{ki} \, , \\
\label{def:Lpar}
L_\parallel [f_{ki}] & = & - \frac{v_{Ti}}{2} \left ( \partial_z \ln{F_0} \, \partial_{v_\parallel} h_{ki} - \partial_{v_\parallel} \ln{F_0} \, \partial_z h_{ki} \right ) \, .
\end{eqnarray}
Here, $h_{ki} = f_{ki} + Z_i F_{0i} J_{0k} \phi_k T_{e0} / T_{i0}$ is the nonadiabatic part of the distribution function, with the ions charge number $Z_i$ and the ion thermal velocity $v_{Ti}$. $T_{i0}$ and $T_{e0}$ are, respectively, the ion and electron equilibrium temperature, $J_{0k}$ is the zeroth order Bessel function corresponding to Fourier transformed gyroaverage operator, and $\phi_k$ is the electrostatic potential. The two terms $K_x$ and $K_y$ are due to magnetic field curvature and gradient introduced by the magnetic geometry\cite{lapillonne-PoP-2009}.

$N$ is the nonlinear term describing the perpendicular advection of the distribution function by the $\mathbf{E} \times \mathbf{B}$ drift velocity:

\begin{equation}
N [\phi_{k}, f_{ki}] = - \sum_{k'_{x,y}} (k_x' k_y - k_x k_y') J_{0{\bf k'}} \phi_{\bf k'} f_{({\bf k}-{\bf k}') i} \, ,
\end{equation}
which has the fundamental role of coupling different perpendicular $k_x$ and $k_y$ modes.

Numerical dissipation terms in GENE have the general form:

\begin{equation}
D [f_{ki}] = a_x k_x^n f_{ki} + a_y k_y^n f_{ki} + a_z \partial_z^4 f_{ki} + a_{v_\parallel} \partial_{v_\parallel}^4 f_{ki} \, ,
\end{equation} where the coefficients $a_x$ and $a_y$ are usually set to zero, while $a_z = 0.1$ and $a_{v_\parallel} = 1$ have been shown to be well adapted in a wide range of cases~\cite{pueschel-CPC-2010}.

The electrostatic potential $\phi_k$ is given by the quasi neutrality equation:

\begin{equation}
\phi_k - \left < \phi_k \right >_{\textrm{\tiny FS}}  + \frac{Z_i T_{e0}}{T_{i0}} \left [ 1 - \Gamma_0 \left ( b_i \right ) \right ] \phi_k = \pi B_0 \int dv_\parallel d\mu J_{0k} f_k \, , 
\label{eq:Poisson}
\end{equation} where $\left < \phi_k \right >_{\textrm{\tiny FS}} = \left ( \int J dz \phi_k \right ) / \left ( \int J dz \right )$,  stands for the flux surface average of the electrostatic potential, $\Gamma_0 (b_i)$ is the modified Bessel function applied to the argument $b_i = v_{Ti}^2 k_\bot^2 / \Omega_{ci}^2$. Electrons are assumed adiabatic: $n_{e} = q_e n_{e0} \left ( \phi_k - \left < \phi_k \right >_{\textrm{\tiny FS}} \right ) / T_{e0}$. Since a single gyrokinetic ion species is considered, the species indices are omitted in the following for the ions distribution function: $f_k = f_{ki}$.

\subsection{Filtered gyrokinetics}

\begin{figure}
\includegraphics[width=8.5cm]{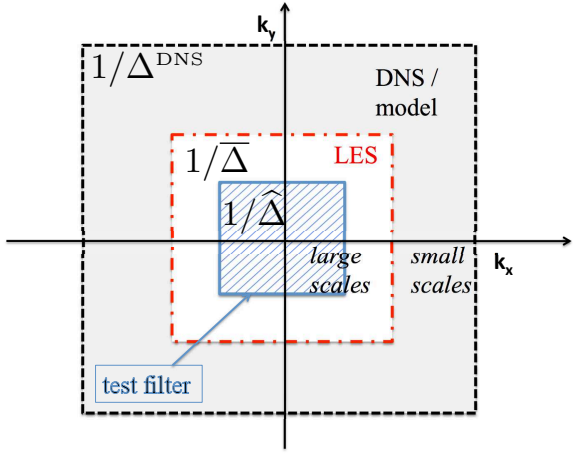}
\caption{Schematic view of a Large Eddy Simulation: The smallest scales (grey area between dashed-dotted and dotted lines) are retained only in a DNS, while they are modeled in a LES model; LES only retain the area inside the dashed-dotted line; alternatively or additionally, a test filter can be used (hatched area, solid line).}
\label{fig:principe-LES}
\end{figure}

In a gyrokinetic LES, the most suitable coordinate subspace for coarsening the grid is the perpendicular wavenumber plane ($k_x$, $k_y$) since it generally requires fairly high resolution. Obviously, the objective of the LES technique is to reduce the number of grid points in ($k_x$, $k_y$) space. The coarsening procedure can be implemented by applying a Fourier low-pass filter, with the characteristic length $\overline{\Delta}$. The employed cut-off filtering has the effect of setting to zero the smallest scales characterized by all modes larger than $1/\overline{\Delta}$, as shown in Fig.~\ref{fig:principe-LES}. If one denotes the action of the filter on the unknowns by $\overline{\cdots}$, the filtered gyrokinetic equation reads:

\begin{equation}
\label{eq:filtered-gyrokinetics}
\partial_t \overline{f_k} = L [\overline{f_k}] + N [\overline{ \phi_k}, \overline{f_k}] + T_{\overline{\Delta}, \Delta^{\textrm{\tiny DNS}}} - D [\overline{f_k}] \, ,
\end{equation} where a new term appears from the filtering of the nonlinear term:

\begin{equation}
T_{\overline{\Delta}, \Delta^{\textrm{\tiny DNS}}} = \overline{N} [\phi_k, f_k] - N [\overline{\phi_k}, \overline{f_k}] \, .
\label{eq:sub-grid}
\end{equation}

At this point, it is important to note that Eq.~(\ref{eq:sub-grid}) is the only term which contains the influence of the scales $\Delta^{\textrm{\tiny DNS}}$ which we want to filter out from ($\phi_k$, $f_k$). Wwe will refer to it as \textit{sub-grid} term in the following. The GyroLES then consists of finding a good model replacing this term which only depends on the resolved unknowns ($\overline{\phi_k}$, $\overline{f_k}$), on the characteristic length of the filter $\overline{\Delta}$, and on some free parameters $\{c_n\}$. 

\subsection{Free energy and sub-grid term}

As has been shown both theoretically \cite{brizard-RMP-2007, schekochihin-PPCF-2008, garbet-NF-2010} and numerically \cite{watanabe-NF-2006, candy-PoP-2006}, the free energy is a relevant quantity for studying gyrokinetic turbulence. The free energy is defined as:

\begin{equation}
\label{def:FE}
\mathcal{E} = \frac{n_{0i} T_{0i}}{V T_{0e}} \sum_{k_x} \sum_{k_y} \int \pi dz dv_\parallel d\mu \frac{h_{-ki} f_{ki}}{2 F_{0i}} \, ,
\end{equation}
with the volume $V = \sum_{k_x} \sum_{k_y} \int dz / B_0$.

The dynamics of the quantity $\mathcal{E}$ can be derived from Eq.~(\ref{eq:gyrokinetic}) by the action of the ``free energy operator'' $\Xi$ on the distribution function $f_{ki}$: $\mathcal{E} = \frac{1}{2} \Xi [ f_{ki} ]$ with
\begin{equation}
\label{def:Xi}
\Xi [\xi_k] = \frac{n_{0i} T_{0i}}{V T_{0e}}\sum_{k_x} \sum_{k_y} \int \pi dz dv_\parallel d\mu \frac{h_{-ki}}{F_{0i}} \xi_k \, .
\end{equation}

One thus obtains:

\begin{equation}
\label{eq:FE-balance}
\partial_t \mathcal{E} = \mathcal{G} - \mathcal{D} \, ,
\end{equation}
with the definitions
\begin{equation}
\label{eq:def-G}
\mathcal{G} = \Xi \left [ L_{G} [f_{ki}] \right ]\,,\quad \mathcal{D} = \Xi \left [ D [f_{ki}] \right ]\,.
\end{equation}

This balance is of particular relevance for the design of a good model. As pointed out in Ref.~\cite{schekochihin-PPCF-2008}, Eq.~(\ref{eq:FE-balance}) involves only quantities which are quadratic in the distribution function, like the kinetic energy in fluid turbulence. Moreover, like the latter quantity, the free energy is injected at large scales by the background gradients and dissipated at various smaller scales by the dissipation terms $\mathcal{D}$. It is important to note in this context that the parallel advection term ($L_\parallel$), the magnetic term ($L_{B_0}$), and the nonlinear term ($N$) have a null contribution to the total free energy balance.

\section{Developing a gyrokinetic LES model \label{sec:model-design}}

As is well known, a naive truncation of small scales can lead to a pile-up of free energy at the smallest scales which are retained in the filtered simulation.\cite{morel-PoP-2011} A good LES model is thus required to dissipate the correct amount of free energy. In the following, the role of sub-grid terms in the free energy balance will be studied in detail. A model will then be developed which agrees as much as possible with the desired sub-grid properties.

\subsection{Sub-grid term and dissipation of free energy}

The nonlinear term has the fundamental role of transferring free energy across perpendicular scales without affecting the global free energy balance. This property is expressed by
\begin{eqnarray}
\label{eq:int-Xi-N}
\Xi \left [ N [\phi_k, f_k] \right ] & = & 0 \, ,
\end{eqnarray}
simply reflecting the fact that the nonlinearity has a Poisson bracket structure and, consequently, its integration cancels upon integration. For the same reason, if a filter is introduced, the following property holds:
\begin{eqnarray}
\label{eq:int-overline-Xi-N}
\overline{\Xi} \left [ N [\overline{\phi_k}, \overline{f_k}] \right ] & = & 0 \, ,
\end{eqnarray}
where $\overline{\Xi}$ is the filtered free energy operator defined in the filtered space. On the contrary, the filtered free energy operator has a non vanishing contribution when it is applied to the sub-grid term:
\begin{eqnarray}
\mathcal{T}_{\overline{\Delta}, \Delta^{\textrm{\tiny DNS}}} & = & \overline{\Xi} [T_{\overline{\Delta}, \Delta^{\textrm{\tiny DNS}}}] = \overline{\Xi} \left [ \overline{N} [\phi_k, f_k] - N [\overline{\phi_k}, \overline{f_k}] \right ] \nonumber \\
& & \hspace{.5cm} = \overline{\Xi} \left [ \overline{N}[\phi_k, f_k] \right ] \, .
\label{eq:TFE}
\end{eqnarray}

The filtered free energy balance can then be expressed as

\begin{equation}
\label{eq:filtered-free-energy}
\partial_t \overline{\mathcal{E}} = \overline{\mathcal{G}} + \mathcal{T}_{\overline{\Delta}, \Delta^{\textrm{\tiny DNS}}} - \overline{\mathcal{D}} \, ,
\end{equation} where filtered quantities are obtained from the action of the filtered free energy operator $\overline{\Xi}$ on the filtered gyrokinetic equation (\ref{eq:filtered-gyrokinetics}).

Recalling that the free energy is assumed to be injected at large scales, then transferred to smaller scales and dissipated there, one can expect that the sub-grid contribution to free energy balance (\ref{eq:TFE}) will be negative. Indeed, one can expect that a large majority of the free energy injection will not be affected by the filtering: $\overline{\mathcal{G}} \approx \mathcal{G}$. It follows that the DNS dissipation can be approximated by $\mathcal{D} \approx \overline{\mathcal{D}} - \mathcal{T}_{\overline{\Delta}, \Delta^{\textrm{\tiny DNS}}}$.

\begin{figure}[h]
\includegraphics[width=8.5cm]{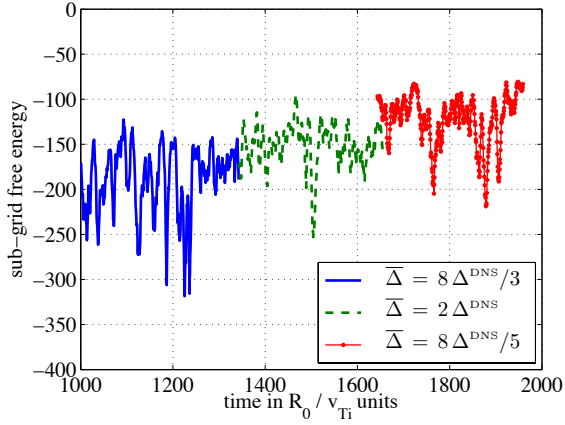}
\caption{Contribution of the sub-grid term to the free energy balance as a function of time, for different test-filter widths $\overline{\Delta}$.}
\label{Fig:sub-grids-filter-width}
\end{figure}

The time evolution of the sub-grid contribution to the filtered free energy balance (CBC for ITG range of parameters) for different values of the filter width $\overline{\Delta}$ is shown in Fig.~\ref{Fig:sub-grids-filter-width}.  One can see that the sub-grid contribution is always negative, implying that the sub-grid scales act as a free energy sink, like it is supposed to.\cite{morel-PoP-2011} More precisely, one observes that the amplitude of the dissipation ensured by the sub-grid scales increases with the filter width. This means that a model should behave like

$$M (c, \overline{\Delta}, \overline{f_k}) = \overline{\Delta}^{\alpha} M' (c, \overline{f_k}) \approx T_{\overline{\Delta}, \Delta^{\textrm{\tiny DNS}}}\,. $$

\subsection{A model for sub-grid scales \label{subsec:def-model}}

A simple dissipative model for GyroLES which has already been used previously \cite{morel-PoP-2011} is given by

\begin{equation}
M(c_\perp, f_{ki}) = c_\perp k_\perp^4 h_{ki} \, .
\end{equation}

The optimal value of $c_\perp$ for the CBC parameters can be found, e.g., through trial and error. However, this model is not taking into account the filter width dependency $\overline{\Delta}$ observed in the previous section. Moreover, the use of $k_\perp$ implies that the relative dissipation in $k_x$ and $k_y$ is fixed. A more flexible model which takes into account the anisotropy ($c_x$ and $c_y$) and the filter width dependency ($\overline{\Delta}_{x, y}$) is given by

\begin{equation}
M = \left ( \overline{\Delta}_x^\alpha c_x k_x^n + \overline{\Delta}_y^\alpha c_y k_y^n \right ) h_{ki}\,.
\end{equation}

In fluid turbulence, it is common to assume that the kinetic energy flux from scale to scale is a constant in the inertial range. Based on the recent finding that ITG turbulence also exhibits a local and direct cascade of free energy \cite{banon-PRL-2011}, we assume, in close analogy, that the free energy flux $\varepsilon_{\mathcal{E}}$ is constant from scale to scale in the ($k_x$, $k_y$) plane perpendicular to the magnetic field. The free energy has the dimension of an energy density, so that the free energy flux $\varepsilon_{\mathcal{E}}$ is an energy density per time,
$$[\varepsilon_{\mathcal{E}}] = \ell^{-1} \tau^{-3} \,, $$
where $\tau$ and $\ell$ represent characteristic time and length scales. It is reasonable to assume that the model depends only on the free energy flux $\varepsilon_{\mathcal{E}}$ and the filter width $\overline{\Delta}$,

$$M = \overline{\Delta}^\alpha \varepsilon_{\mathcal{E}}^\beta k^n h_k \, .$$

Moreover, from dimensional analysis we know that $[M] = \tau^{-1} [h_k]$, so that $\beta = 1/3$ and $\alpha = n + 1/3$. The last relation allows to fix the unknown filter width exponent $\alpha$ accordingly to the model parameter $n$. The model thus becomes

\begin{equation}
M = \left ( c_x \overline{\Delta}_x^{n + 1/3} k_x^n + c_y \overline{\Delta}_y^{n + 1/3} k_y^n \right ) h_{ki}\,.
\label{eq:model-2}
\end{equation}

Since the derivative order $n$ is positive, the filter width exponent $\alpha = n + 1/3$ is also positive, in line with the numerical results in the previous section. Moreover, the model coefficients are dimensionally related to the constant free energy flux across scales via $[c_x] = [c_y] = [\varepsilon_{\mathcal{E}}]^{1/3}$. It is interesting to note here that the model coefficients are constants, just like the free energy flux.

\subsection{Dynamic procedure for gyrokinetics \label{subsec:dynproc}}

The dynamic procedure is based on the introduction of an additional filter denoted by $\widehat{\cdots}$ and referred to as the test-filter. It is characterized by a filter width $\widehat{\Delta}$ that corresponds to a ``very coarse'' grid: $\widehat{\Delta} > \overline{\Delta} > \Delta^{\textrm{\tiny DNS}}$. The gyrokinetic equation associated to the test-filter grid can be obtained by test-filtering the gyrokinetic equation expressed in the DNS domain:

\begin{equation}
\label{eq:gyrokinetic-test}
\partial_t \widehat{f_k} = L[\widehat{f_k}] + N [\widehat{\phi_k}, \widehat{f_k}] - D[\widehat{f_k}] + T_{\widehat{\Delta}, \Delta^{\textrm{\tiny DNS}}} \, .
\end{equation}
This equation is equivalent to the LES filtered Eq.~(\ref{eq:filtered-gyrokinetics}) with the LES width ($\overline{\Delta}$) replaced by the test-filter one ($\widehat{\Delta}$).

Alternatively, the equation in the test-filter domain can be obtained by test-filtering ($\widehat{\Delta}$) the gyrokinetic equation expressed in the LES domain, Eq.~(\ref{eq:filtered-gyrokinetics}),

\begin{equation}
\label{eq:gyrokinetic-test-LES}
\partial_t \widehat{f_k} = L[\widehat{f_k}] + \widehat{N} [\overline{\phi_k}, \overline{f_k}] - D[\widehat{f_k}] + \widehat{T}_{\overline{\Delta}, \Delta^{\textrm{\tiny DNS}}} \,,
\end{equation}
where we have used the very important property $\widehat{\overline{\cdots}} = \widehat{\cdots}$ of Fourier cutoff filters. Comparing Eqs.~(\ref{eq:gyrokinetic-test}) and (\ref{eq:gyrokinetic-test-LES}), one obtains the Germano identity,

\begin{eqnarray}
T_{\widehat{\Delta}, \Delta^{\textrm{\tiny DNS}}} & = & \widehat{T}_{\overline{\Delta}, \Delta^{\textrm{\tiny DNS}}} + \widehat{N} [\overline{\phi_k}, \overline{f_k}] - N [\widehat{\phi_k}, \widehat{f_k}] \, , \nonumber \\
& = & \widehat{T}_{\overline{\Delta}, \Delta^{\textrm{\tiny DNS}}} + T_{\widehat{\Delta}, \overline{\Delta}} \, .
\label{eq:germano}
\end{eqnarray}

During an LES, the sub-grid term $T_{\widehat{\Delta}, \overline{\Delta}}$ can be computed exactly, since it involves test filtering ($\widehat{\Delta}$) of the LES-resolved quantities ($\overline{\Delta}$). On the other hand, the two other terms involve the non-resolved DNS scales ($\Delta^{\textrm{\tiny DNS}}$) and therefore have to be approximated by the model: 

\begin{equation}
\label{eq:sub-grids-approx}
T_{\widehat{\Delta}, \Delta^{\textrm{\tiny DNS}}} \approx M_{\widehat{\Delta}} \,\,\,\,\, ; \hspace{0.5cm} T_{\overline{\Delta}, \Delta^{\textrm{\tiny DNS}}} \approx M_{\overline{\Delta}} \, .
\end{equation}

The dynamic procedure consists of introducing the model approximations, Eq.~(\ref{eq:sub-grids-approx}), into the Germano identity, Eq.~(\ref{eq:germano}), to obtain

\begin{equation}
\label{eq:approx-germano}
M_{\widehat{\Delta}} \approx \widehat{M}_{\overline{\Delta}} + T_{\widehat{\Delta}, \overline{\Delta}} \, .
\end{equation}

Since the model is an approximation of the sub-grid term, Eq.~(\ref{eq:germano}) can only be approximated during an LES. Now, one can define the squared distance $d^2$ which is to minimize via

\begin{equation}
\label{def:distance}
d^2 = \left < \left ( T_{\widehat{\Delta}, \overline{\Delta}} + \widehat{M}_{\overline{\Delta}} - M_{\widehat{\Delta}} \right )^2 \right >_\Lambda \, ,
\end{equation}
where $\left < \cdots \right >_\Lambda$ stand for integration over the entire phase space. 

As was shown in Sec.~\ref{subsec:def-model}, the model coefficients $c_x$ and $c_y$ can be assumed to be constant in the gyrokinetic ``inertial range.'' So provided that the range between test-filter and LES scales belongs to this ``inertial range,'' the coefficients do not depend on the filter widths ($\widehat{\Delta}$, $\overline{\Delta}$).

Using Eq.~(\ref{eq:model-2}), the squared distance can be expressed in terms of the model amplitudes $c_x$ and $c_y$ according to
\begin{equation}
\label{eq:distance}
d^2 = \left < \left ( T_{\widehat{\Delta}, \overline{\Delta}} + c_x m_x + c_y m_y \right )^2 \right >_\Lambda \, ,
\end{equation}
where the notations $m_{x, y} = \left ( \overline{\Delta}_{x, y}^\alpha - \widehat{\Delta}_{x, y}^\alpha \right ) k_{x, y}^n \widehat{h}_k$ have been introduced.

An optimization of this difference with respect to the unknown parameters ($\partial d^2/\partial c_x = 0$ and $\partial d^2/\partial c_y = 0$) leads to the expressions

\begin{eqnarray}
\label{def:c_x}
c_x & = & \frac{\left < m_x T_{\widehat{\Delta}, \overline{\Delta}} \right >_\Lambda \left < m_y^2 \right >_\Lambda - \left < m_y T_{\widehat{\Delta}, \overline{\Delta}} \right >_\Lambda \left < m_y m_x \right >_\Lambda}{\left < m_x m_y \right >_\Lambda^2 - \left < m_x^2 \right >_\Lambda \left < m_y^2 \right >_\Lambda}\\
\label{def:c_y}
c_y & = & \frac{\left < m_y T_{\widehat{\Delta}, \overline{\Delta}} \right >_\Lambda \left < m_x^2 \right >_\Lambda - \left < m_x T_{\widehat{\Delta}, \overline{\Delta}} \right >_\Lambda \left < m_y m_x \right >_\Lambda}{\left < m_x m_y \right >_\Lambda^2 - \left < m_x^2 \right >_\Lambda \left < m_y^2 \right >_\Lambda} \, .
\end{eqnarray}

Thus, these two free parameters of the model can be computed dynamically during a numerical simulation from Eqs.~(\ref{def:c_x}) and (\ref{def:c_y}). Since they are always positive, it is guaranteed that the model has a dissipative effect on the free energy.

\section{Numerical results \label{sec:numerics}}

In the following, we will present numerical results obtained by means of the dynamic procedure with the GENE code. The set of parameters corresponds to the Cyclone Base Case commonly used for studying Ion Temperature Gradient (ITG) driven turbulence \cite{dimits-PoP-2000}. Considering a minor radius $r_0/R_0 = 0.18$, the density and temperature gradients are, respectively, $\omega_{ni} = 2.22$ and $\omega_{Ti} = 6.96$, the magnetic configuration is characterized by the safety factor $q = 1.4$ and the magnetic shear $\hat{s} = 0.796$, with ions and electrons such that $T_{e0}/T_{i0}=1$ and $Z_i = 1$.

\subsection{Nonlinear Gyrokinetic Large Eddy Simulation: Cyclone Base Case}

For the reference DNS, a perpendicular grid of $N_x \times N_y=128\times 64$ is used. This grid has been used both with and without a LES model, and the results obtained have not been affected, indicating that the simulation is well resolved. On the other hand, a minimal perpendicular grid for GyroLES should be $N_x\times N_y = 48\times 32$, allowing the dynamic procedure to work. Indeed, the use of the latter involves the introduction (in the LES domain $\overline{\Delta}$) of a test filter corresponding to a coarser grid, $\widehat{\Delta} > \overline{\Delta}$. However, it is necessary for the dynamic procedure that the domain of the LES grid which is neglected by the test filter belongs to the gyrokinetic "inertial" range, so that the model coefficients have the same values in the two domains. Here, we will employ a test filter width which corresponds to the half of the LES domain: $\widehat{\Delta}_x = 2 \overline{\Delta}_x$, $\widehat{\Delta}_y = 2 \overline{\Delta}_x$. This means that the optimization in the dynamic procedure is related to a sub-grid term $T_{\widehat{\Delta}, \overline{\Delta}}$ defined by 24 $k_x$ modes and 16 $k_y$ modes. The parameters given in Eqs.~(\ref{def:c_x}) and (\ref{def:c_y}) are computed at each time step of the simulation. The parallel and velocity grids are kept fixed at $N_z = 32$, $N_{v_\parallel} = 64$, and $N_{\mu} = 8$. The model order is chosen to be $n = 4$, leading to
\begin{equation}
\textrm{M4} = \left ( c_x \overline{\Delta}^{13/3}_x k_x^4 + c_y \overline{\Delta}^{13/3}_y k_y^4 \right ) h_{ki} \, .
\label{def:M4}
\end{equation}
We note in passing that the case $n = 2$ has also been tested; in that case, the obtained results showed a high volatility, though. A similar result has been obtained by Smith and Hammett\cite{smith-PoP-1997}, where hyper-viscosity models were found to perform better than viscosity models for gyrofluid turbulence. For comparison, we also show simulations without any model (LES M0), with a perpendicular grid identical to the LES M4 one.

\begin{figure}
\includegraphics[width=8.5cm]{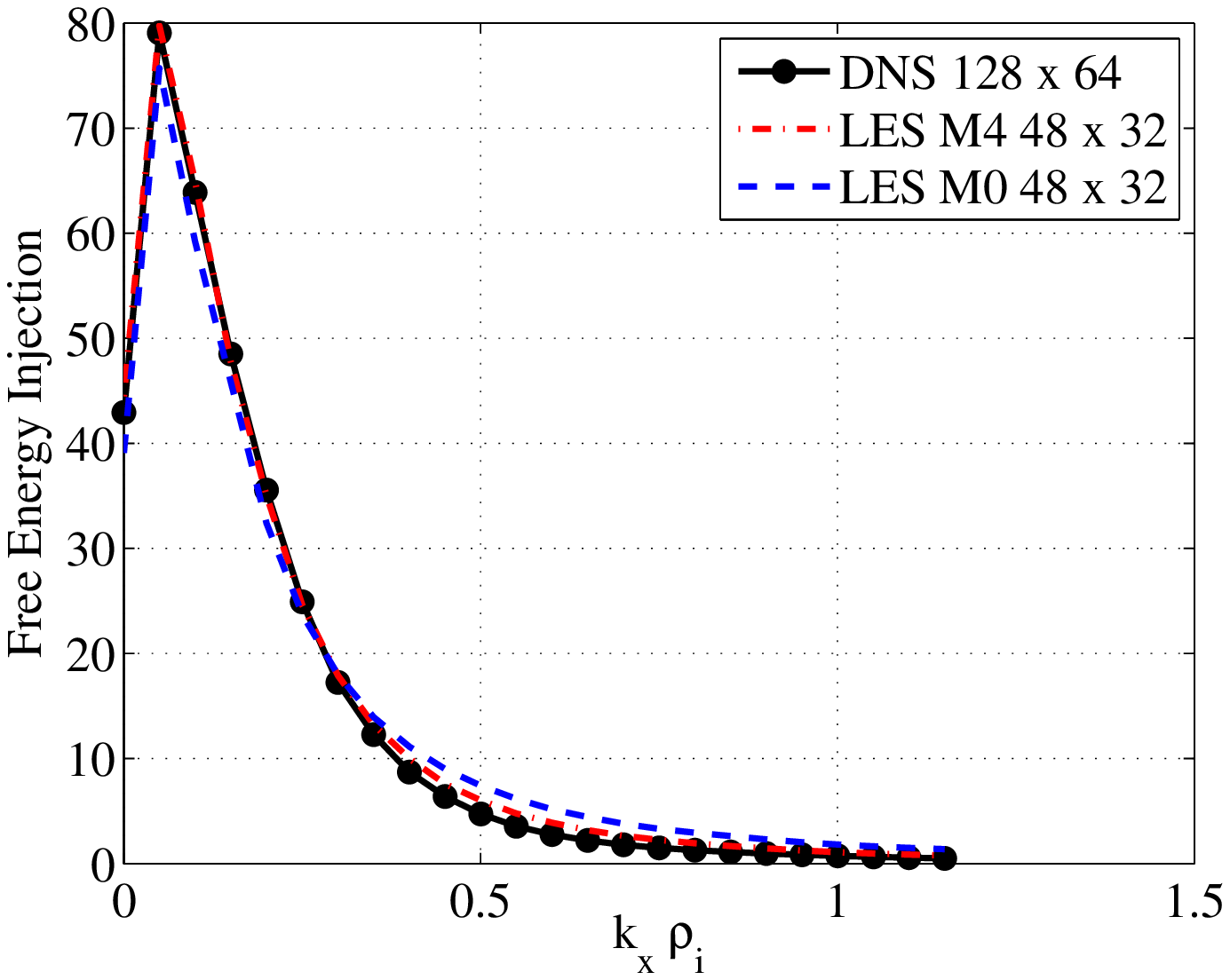}\\
\includegraphics[width=8.5cm]{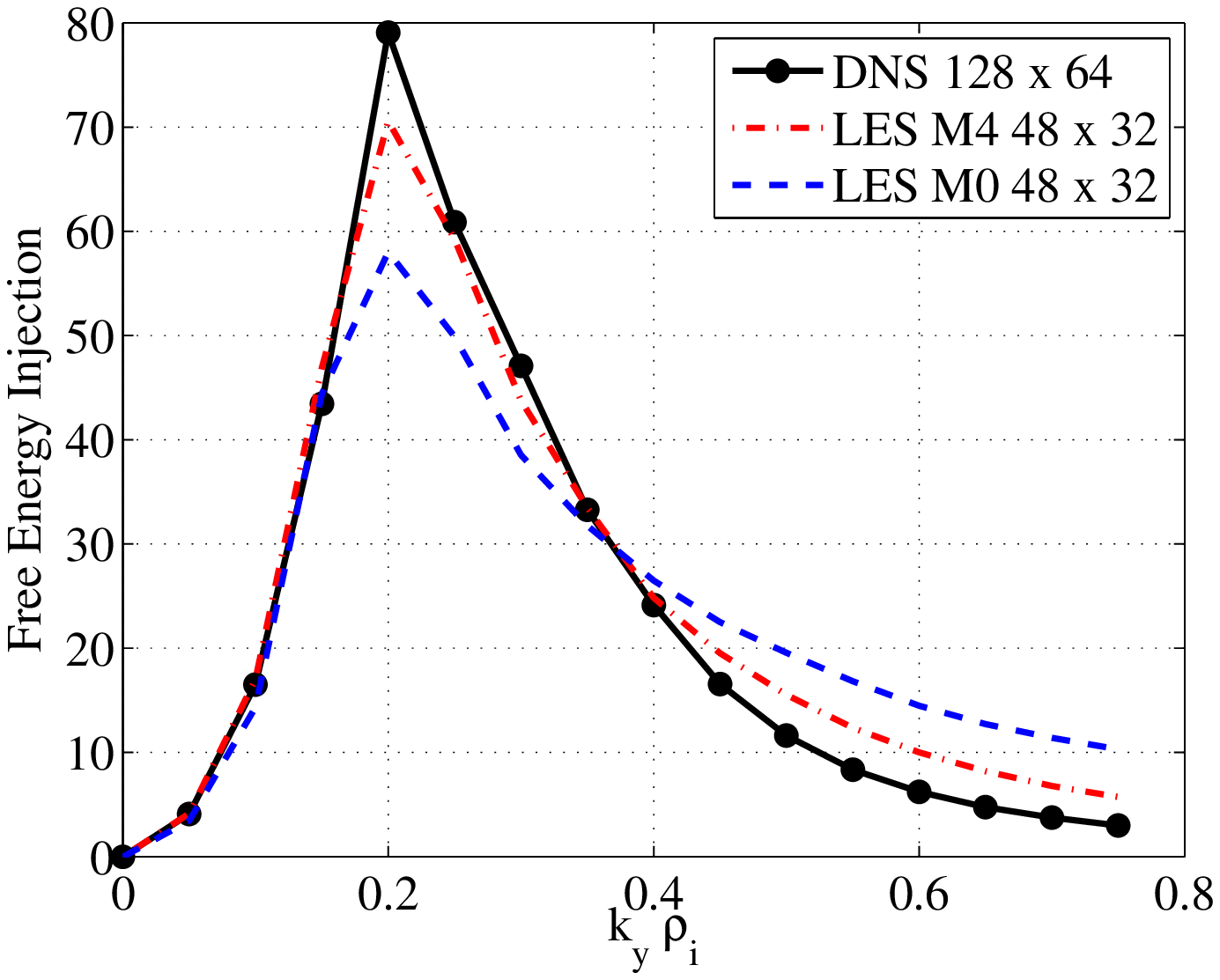}
\caption{Free energy injection spectra ($\mathcal{G}^{k_x}$ at top, $\mathcal{G}^{k_y}$ at bottom) for the fourth-order model (M4) at reduced resolution, compared with a highly resolved DNS and the case without a model (M0).}
\label{fig:cmp_Gf_CBC}
\end{figure}

Since it is proportional to the total heat flux $\mathcal{Q}$, the free energy injection term $\mathcal{G} = \omega_{Ti} \mathcal{Q}$ is of special relevance for comparisons with experimental results and earlier works. The comparisons are based on two dimensional wavenumber spectra of the free energy injection rate,

\begin{equation}
\mathcal{G}^{k_x, k_y} = \frac{n_{0i} T_{i0}}{ V T_{e0}} \int \pi dz dv_\parallel d\mu \left ( \frac{h_{-ki}}{2 F_{0i}} L_G[f_{ki}]\right )\,.
\label{eq:2D_Gkxky}
\end{equation}
It is understood that this quantity is averaged during the quasistationary turbulent state over sufficiently long time windows (at least 2000 $R_0/v_{Ti}$). The reduction to a one-dimensional spectrum is then simply provided by

\begin{equation}
\mathcal{G}^{k_x} = \sum_{k_y} \mathcal{G}^{k_x, k_y} \, , \,\,\,\,\, \mathcal{G}^{k_y} = \sum_{k_x} \mathcal{G}^{k_x, k_y} \, .
\label{eq:1D_FE}
\end{equation}

\begin{figure}
\includegraphics[width=8.5cm]{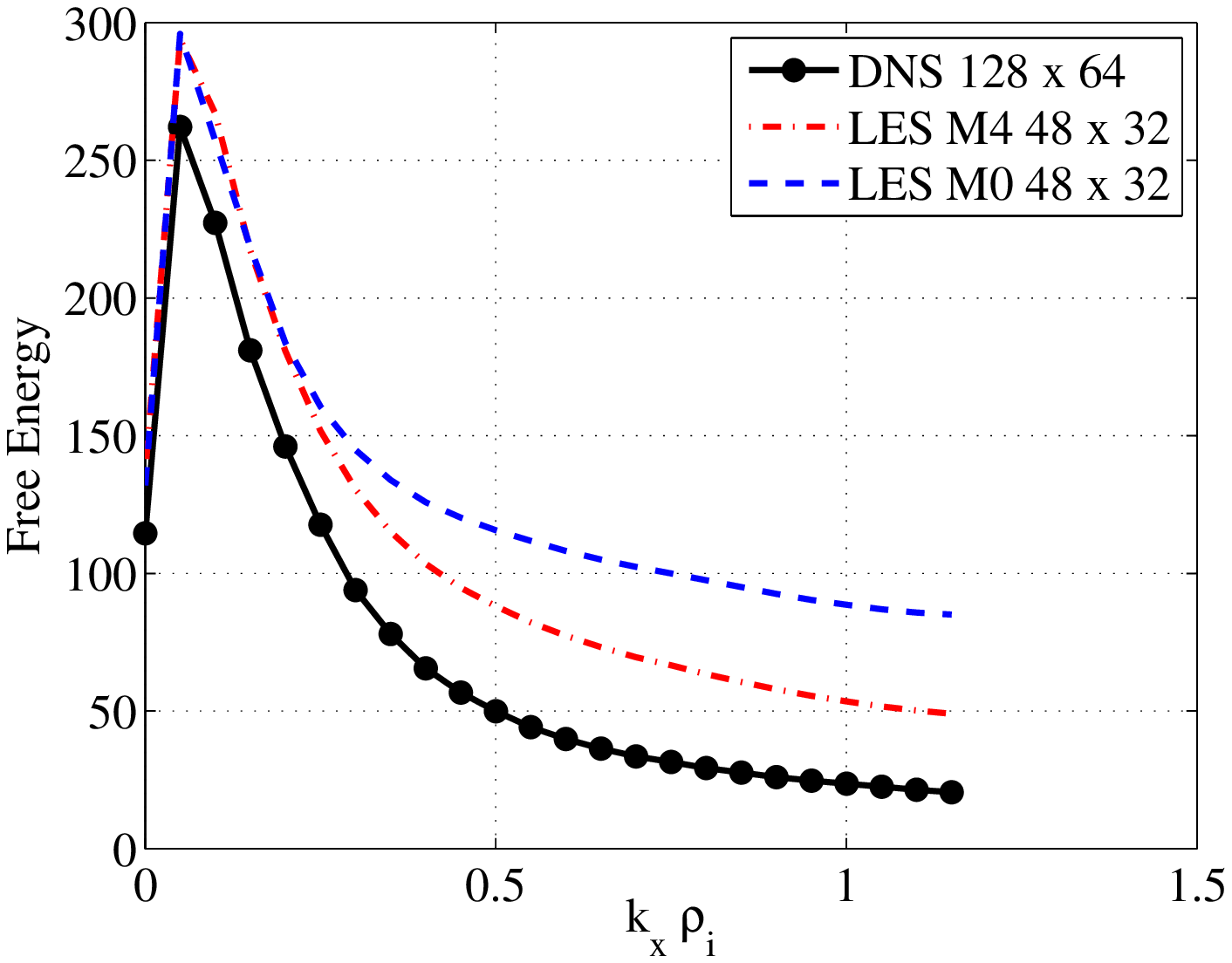}\\
\includegraphics[width=8.5cm]{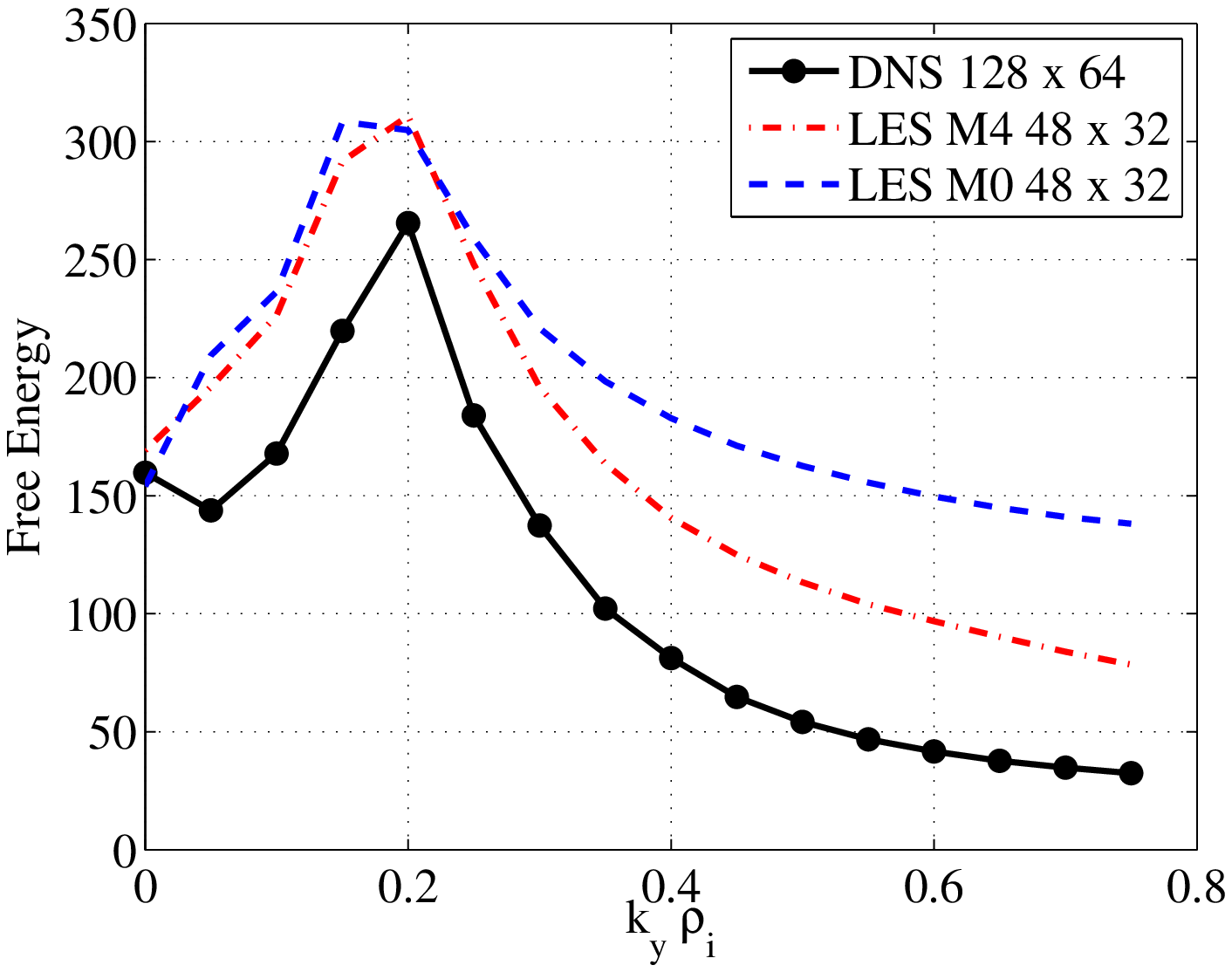}
\caption{Free energy spectra ($\mathcal{E}^{k_x}$ at top, $\mathcal{E}^{k_y}$ at bottom) for the fourth-order model (M4) at reduced resolution, compared with a highly resolved DNS and the case without a model (M0).}
\label{fig:cmp_FE_CBC}
\end{figure}

Comparisons of free energy injection spectra from DNS, GyroLES, and a simulation without a model are shown in Fig.~\ref{fig:cmp_Gf_CBC}. The $k_x$ spectra for all three cases are found to exhibit a surprisingly good level of agreement, but the $k_y$ spectra illustrate that the use of a LES model diminishes the accumulation at the smallest scales, improving the agreement at the largest scales with the reference DNS.

One-dimensional free energy spectra, $\mathcal{E}^{k_x}$ and $\mathcal{E}^{k_y}$, can be constructed in analogy with Eq.~(\ref{eq:1D_FE}). As can be observed in Fig.~\ref{fig:cmp_FE_CBC}, the GyroLES clearly prevents the accumulation of free energy at the smallest scales. However, there still exists an overestimation of the free energy at the largest scales when compared with the reference high-resolution DNS.

Since the LES spectra are truncated with respect to the DNS ones, estimates for the neglected parts of the spectra are required for computing total (integral) values of the heat flux and the free energy. Such estimates may be provided via a power law regression of the spectra.\cite{morel-PoP-2011} The estimate from the GyroLES run yields $\mathcal{E}^{\textrm{\tiny M4}} = 1.49 \, \mathcal{E}^{\textrm {\tiny DNS}}$, while for the case without a model (M0), one obtains $\mathcal{E}^{\textrm {\tiny M0}} = 2.36 \mathcal{E}^{\textrm {\tiny DNS}}$. The total heat flux levels from DNS and LES are in very good agreement, $\mathcal{Q}^{\textrm{\tiny M4}} = 1.06 \, \mathcal{Q}^{\textrm{\tiny DNS}}$. In the case without a model, one finds $\mathcal{Q}^{\textrm{\tiny M0}} = 1.20 \, \mathcal{Q}^{\textrm{\tiny DNS}}$ due to an overestimate at the smallest scales.

For the sake of clarity, all further comparisons will focus on the spectra $\mathcal{E}^{k_y}$ and $\mathcal{G}^{k_y}$ which have been found to be most sensitive. The LES model uses $n=4$ and filter widths such that $\widehat{\Delta} = 2 \overline{\Delta}$; the perpendicular grid size is $N_x\times N_y = 48\times 32$.

\subsection{Robustness while varying the temperature gradient}

As is well known, the logarithmic temperature gradient $\omega_{Ti}$ is a key parameter for ITG turbulence, given that the equilibrium temperature profile acts as a source of free energy for the system. In the following, the robustness of the LES approach is tested for two values of the temperature gradient which differ from the nominal value; these correspond to a weakly driven turbulence case ($\omega_{Ti} = 6.0$) and to a strongly driven turbulence case ($\omega_{Ti} = 8.0$).

\begin{figure}
\includegraphics[width=8.5cm]{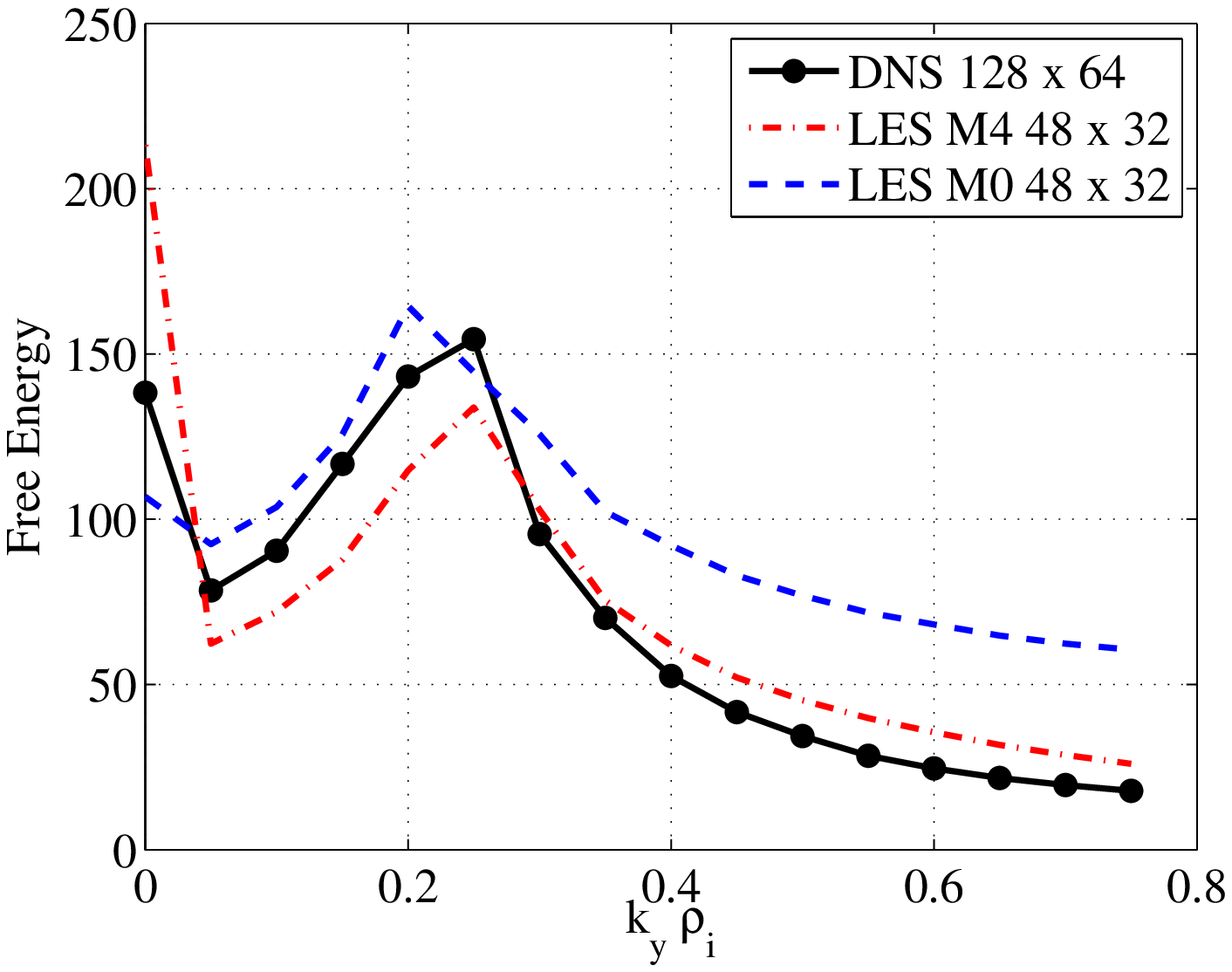}\\
\includegraphics[width=8.5cm]{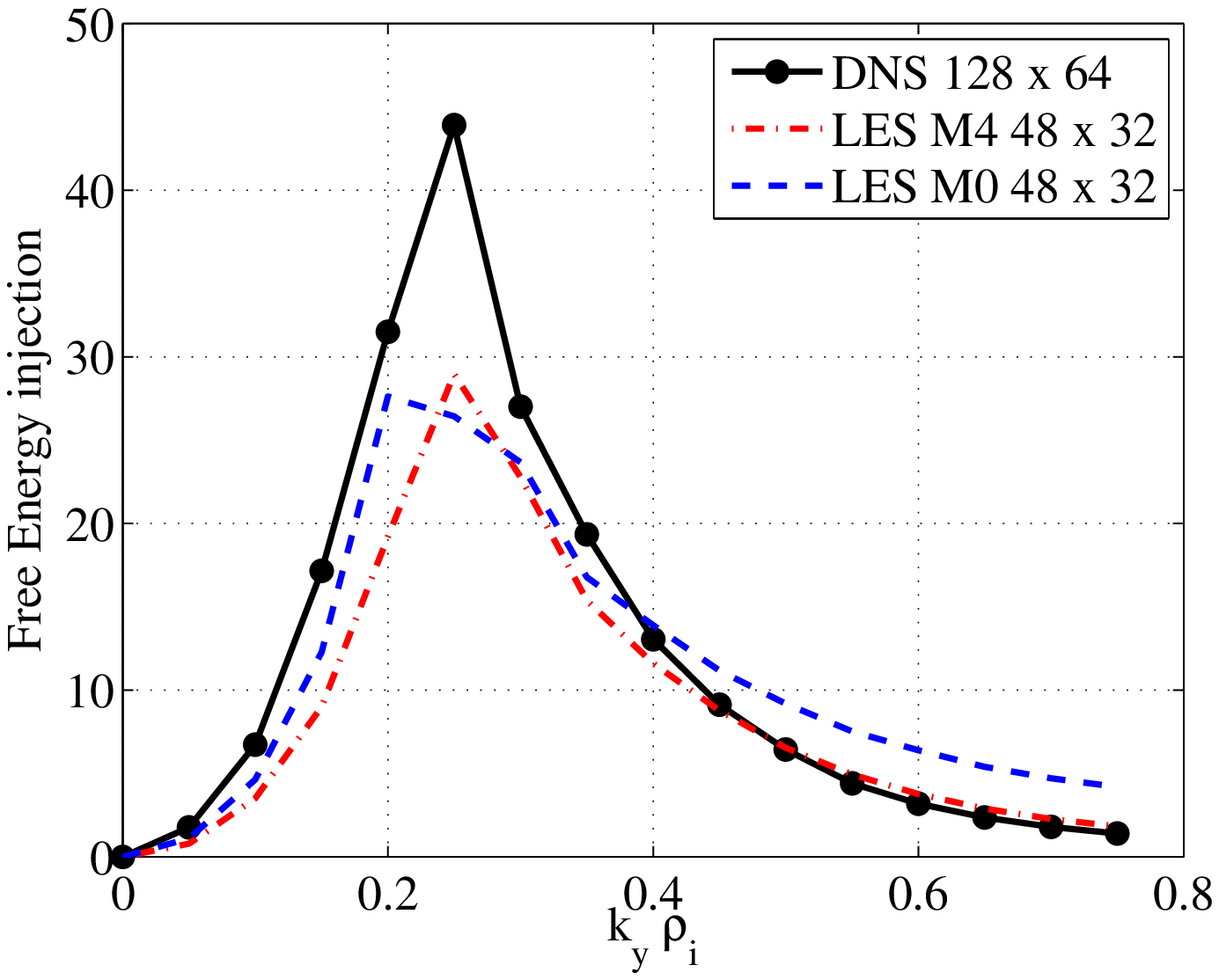}
\caption{Wavenumber spectra $\mathcal{E}^{k_y}$ (at top) and $\mathcal{G}^{k_y}$ (at bottom): Comparison between DNS and LES for the case of weakly driven ITG turbulence at $\omega_{Ti} = 6.0$.}
\label{fig:cmp_LT6}
\end{figure}

The case of weakly driven ITG turbulence is shown in Fig.~\ref{fig:cmp_LT6}. The M4 model yields a very reasonable agreement with the DNS regarding both the free energy spectrum $\mathcal{E}^{k_y}$ and the free energy injection spectrum $\mathcal{G}^{k_y}$. The total values $\mathcal{E}^{\textrm{\tiny M4}} = 1.02 \, \mathcal{E}^{\textrm {\tiny DNS}}$ and $\mathcal{Q}^{\textrm{\tiny M4}} = 1.25 \, \mathcal{Q}^{\textrm {\tiny DNS}}$ are also in good agreement. Without a model, one obtains $\mathcal{E}^{M_0} = 1.79 \, \mathcal{E}^{\textrm {\tiny DNS}}$ and $\mathcal{Q}^{\textrm{\tiny M0}} = 1.04 \, \mathcal{Q}^{\textrm {\tiny DNS}}$. The latter result is accidental, however, and results from a compensation between an underestimation at large scales and an overestimation at small ones.

\begin{figure}
\includegraphics[width=8.5cm]{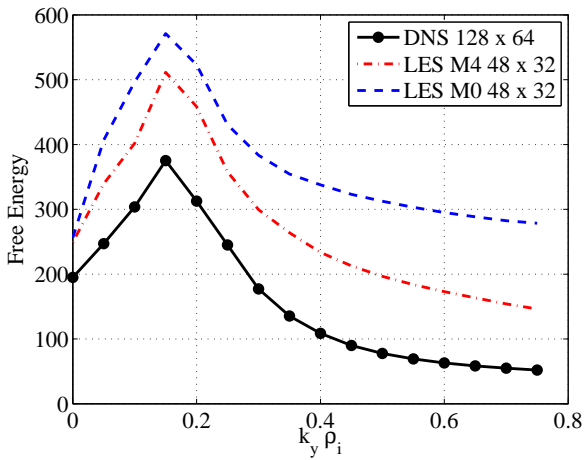}\\
\includegraphics[width=8.5cm]{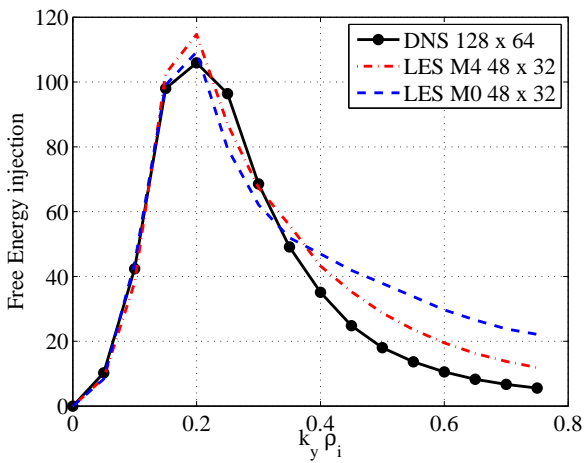}
\caption{Wavenumber spectra $\mathcal{E}^{k_y}$ (at top) and $\mathcal{G}^{k_y}$ (at bottom): Comparison between DNS and LES for the case of strongly driven ITG turbulence at $\omega_{Ti} = 8.0$.}
\label{fig:cmp_FE_ky_LT8}
\end{figure}

Fig.~\ref{fig:cmp_FE_ky_LT8} displays the results for the case of strongly driven ITG turbulence. The LES is found to systematically overestimate the DNS free energy spectrum $\mathcal{E}^{k_y}$, while the prediction of the free energy injection spectrum $\mathcal{G}^{k_y}$ is in reasonable agreement. One finds $\mathcal{E}^{\textrm{\tiny M4}} = 1.67 \, \mathcal{E}^{\textrm {\tiny DNS}}$ and $\mathcal{Q}^{\textrm{\tiny M4}} = 1.14 \, \mathcal{Q}^{\textrm {\tiny DNS}}$, whereas the values exhibit a substantial disagreement without a model, according to $\mathcal{E}^{\textrm{\tiny M0}} = 3.00 \, \mathcal{E}^{\textrm {\tiny DNS}}$ and $\mathcal{Q}^{\textrm{\tiny M0}} = 1.42 \, \mathcal{Q}^{\textrm {\tiny DNS}}$.

In summary, the LES model leads to a far better agreement with the reference DNS than the runs without a model. As far as the overall heat flux levels (which are of prime importance) are concerned, the relative error with respect to the reference DNS is acceptable, amounting to less than $30\%$ in all three cases considered. The model amplitudes $c_x$ and $c_y$ computed dynamically are found to be quite robust when varying the temperature gradient. The mean values are $c_x = 0.0155$, $c_y = 0.0179$ in the weakly driven case, $c_x = 0.0140$, $c_y = 0.0212$ for the CBC, and $c_x = 0.0140$, $c_y = 0.0219$ for the strongly driven case.

\subsection{Robustness while varying the magnetic shear}

Next, we would like to investigate the robustness of the LES approach with respect to variations of the magnetic shear $\hat{s}$. The effects of the latter on plasma microturbulence has been the subject of numerous experimental \cite{fujita-PRL-1997, greenfield-PoP-1997, synakowski-PoP-1997, gormezano-PRL-1998, wolf-PoP-2000}, theoretical \cite{antonsen-PoP-1996}, as well as numerical \cite{kinsey-PoP-2006, garbet-PoP-2001, waltz-PoP-1995, beer-PoP-1997} studies. In this context, it was also found that negative magnetic shear can help improve the plasma confinement in a tokamak by decreasing the level of turbulence. Apart from the CBC case, three highly resolved DNS runs have been performed, corresponding to reversed ($\hat{s} = -0.4$), low ($\hat{s} = 0.2$), or high ($\hat{s} = 1.4$) magnetic shear cases, compared to the CBC standard value ($\hat{s} = 0.796$). The DNS perpendicular grid is kept fixed compared to previous sections: $N_x \times N_y = 128\times 64$, while other parameters are those of the CBC.

\begin{figure}
\includegraphics[width=8.5cm]{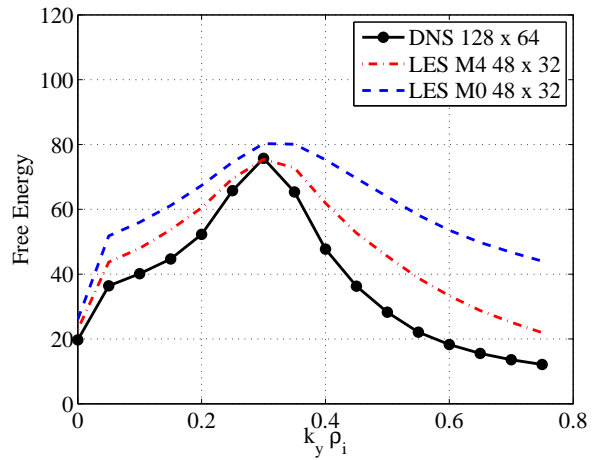}\\
\includegraphics[width=8.5cm]{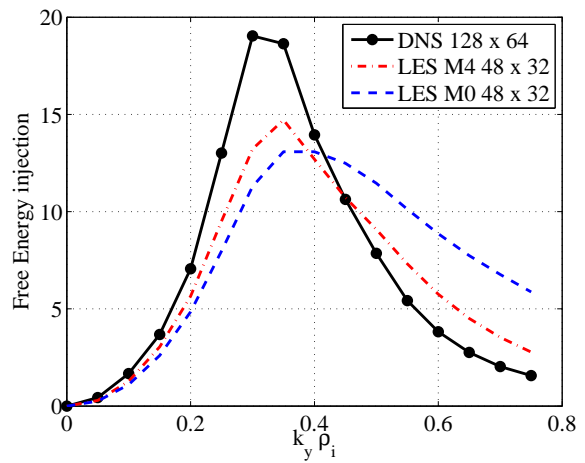}
\caption{Wavenumber spectra $\mathcal{E}^{k_y}$ (at top) and $\mathcal{G}^{k_y}$ (at bottom): Comparison between DNS and LES for the case of reversed shear ($\hat{s} = -0.4$).}
\label{fig:cmp_FE_ky_s-04}
\end{figure}

In the case of reversed shear, the free energy and free energy injection spectra peak at a slightly higher $k_y$ value ($k_y \rho_i \simeq 0.3$ compared to $k_y \rho_i \simeq 0.2$ for CBC), as shown in Fig.~\ref{fig:cmp_FE_ky_s-04}. The total free energy is very small compared to the CBC, indicating a low level of turbulence. This effect has already been observed in a previous numerical study based on the spectral heat flux \cite{kinsey-PoP-2006}. The LES offers a satisfying agreement with the reference DNS spectra, except for an underestimation of the free energy injection peak. The total free energy agrees reasonably well with the reference value, $\mathcal{E}^{\textrm{\tiny M4}} = 1.35 \, \mathcal{E}^{\textrm {\tiny DNS}}$, while $\mathcal{E}^{\textrm {\tiny M0}} = 2.12 \, \mathcal{E}^{\textrm {\tiny DNS}}$. Considering the heat fluxes, one finds the same trend: $\mathcal{Q}^{\textrm{\tiny M4}} = 1.04 \, \mathcal{Q}^{\textrm {\tiny DNS}}$ and $\mathcal{Q}^{\textrm {\tiny M0}} = 1.31 \, \mathcal{Q}^{\textrm {\tiny DNS}}$.

\begin{figure}
\includegraphics[width=8.5cm]{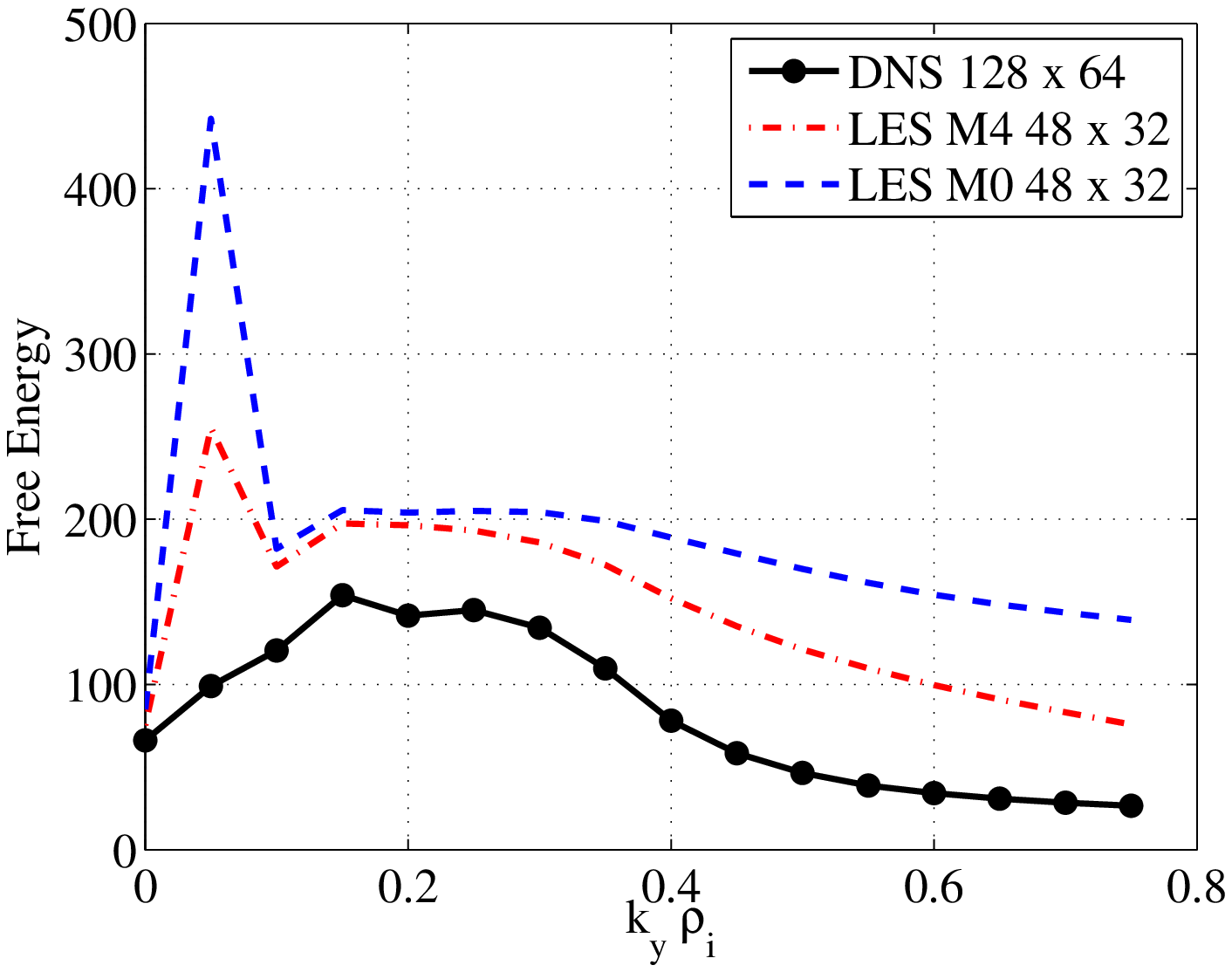}\\
\includegraphics[width=8.5cm]{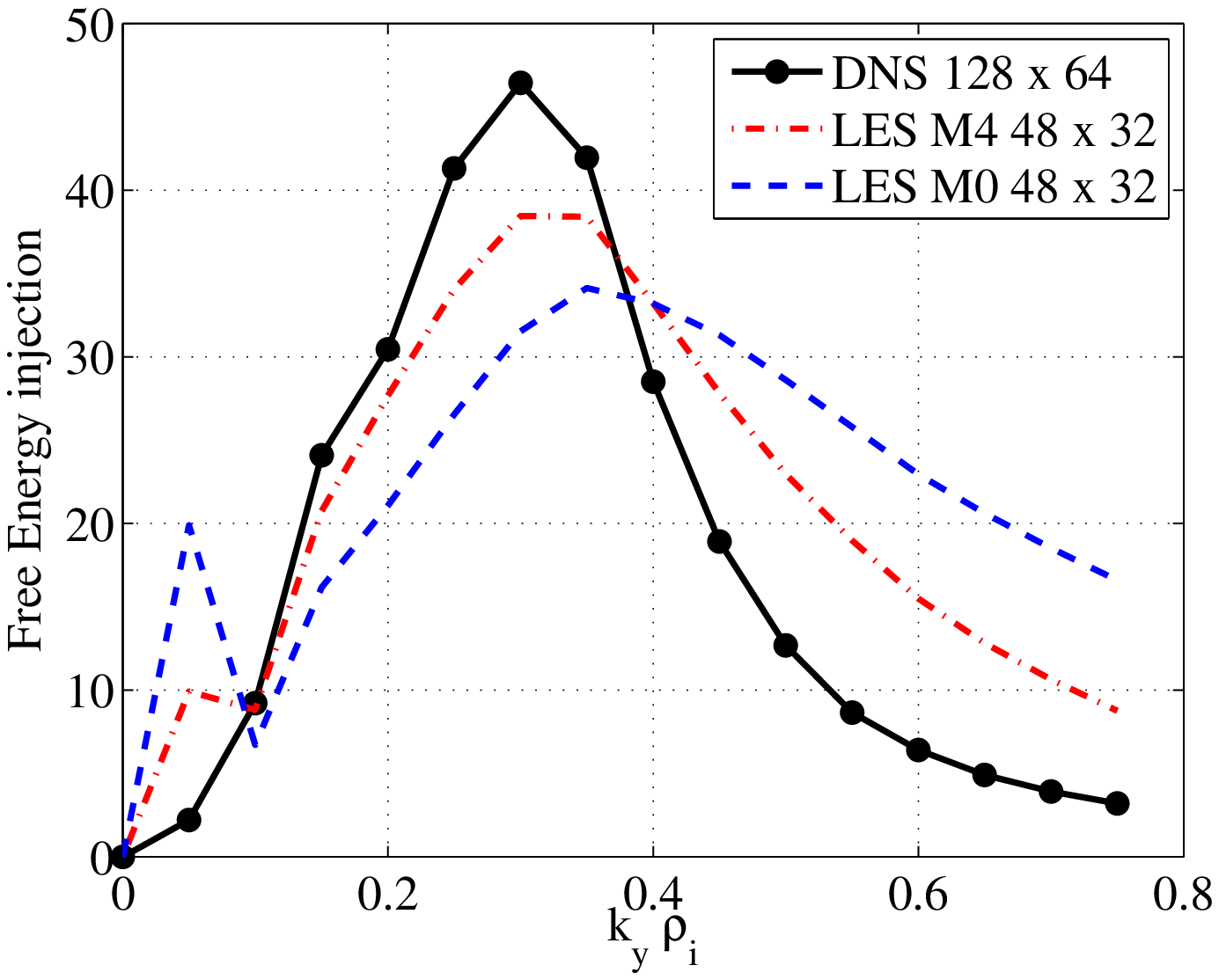}
\caption{Wavenumber spectra $\mathcal{E}^{k_y}$ (at top) and $\mathcal{G}^{k_y}$ (at bottom): Comparison between DNS and LES for the case of low shear ($\hat{s} = 0.2$).}
\label{fig:cmp_FE_ky_s02}
\end{figure}

In Fig.~\ref{fig:cmp_FE_ky_s02}), the results for the case of low shear ($\hat{s} = 0.2$) are presented. Here, the turbulence level lies between those of the reversed shear and CBC cases. The free energy and free energy injection spectra, $\mathcal{E}^{k_y}$ and $\mathcal{G}^{k_y}$, are fairly extended, up to $k_y \rho_i \approx 0.4$. The use of a LES model prevents the accumulation of free energy at small scales, while it moderates the appearance of large-scale structures without suppressing them completely (at small non-zero $k_y$). The total free energy obtained by the LES model is a bit far from the reference DNS value, $\mathcal{E}^{\textrm{\tiny M4}} = 1.78 \, \mathcal{E}^{\textrm {\tiny DNS}}$, but much better than the estimate obtained without a model, $\mathcal{E}^{\textrm {\tiny M0}} = 2.89 \, \mathcal{E}^{\textrm {\tiny DNS}}$. The disagreement regarding the heat fluxes is again found to be more acceptable, according to $\mathcal{Q}^{\textrm{\tiny M4}} = 1.17 \, \mathcal{Q}^{\textrm {\tiny DNS}}$ and $\mathcal{Q}^{\textrm{\tiny M0}} = 1.56 \, \mathcal{Q}^{\textrm {\tiny DNS}}$.

\begin{figure}
\includegraphics[width=8.5cm]{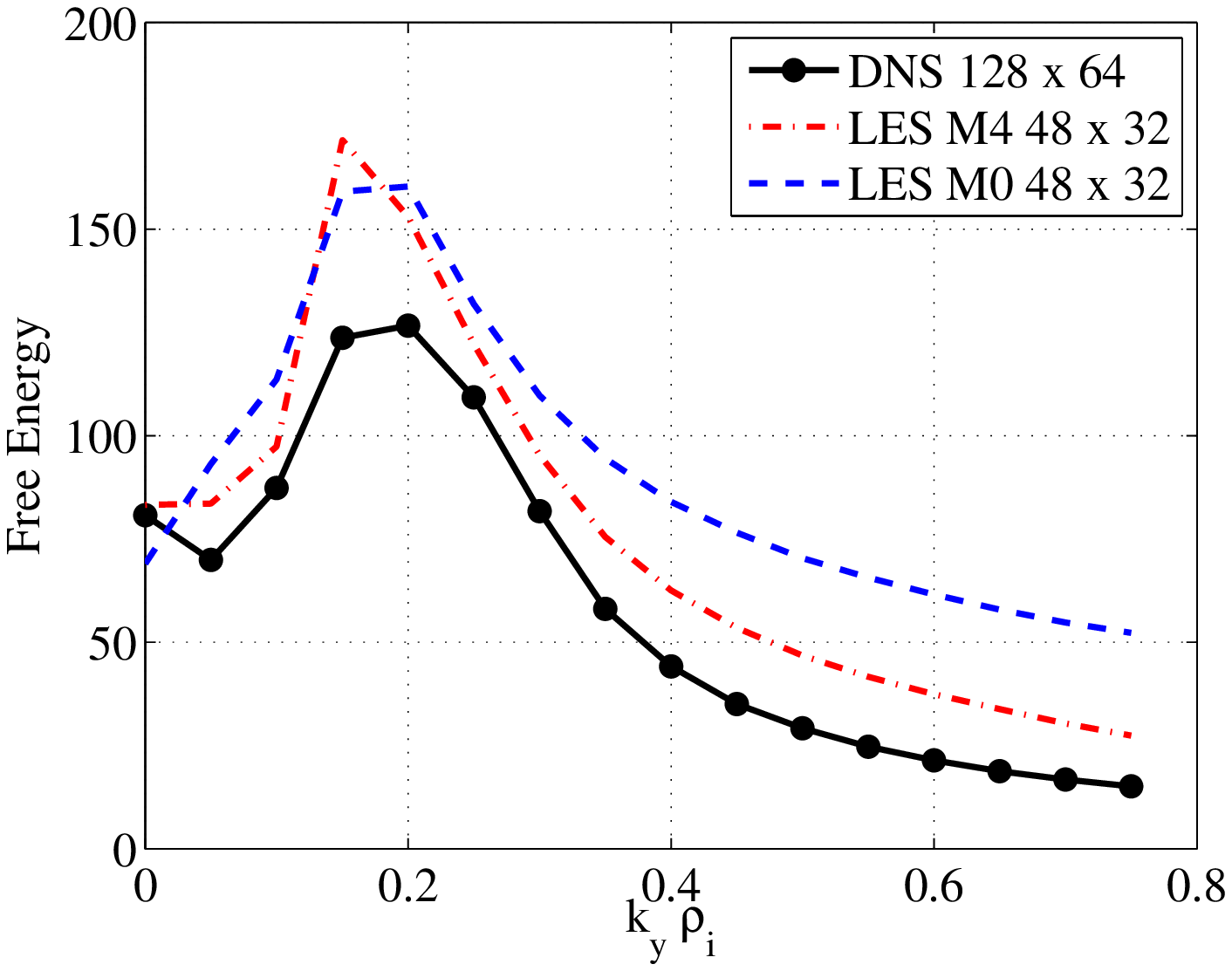}\\
\includegraphics[width=8.5cm]{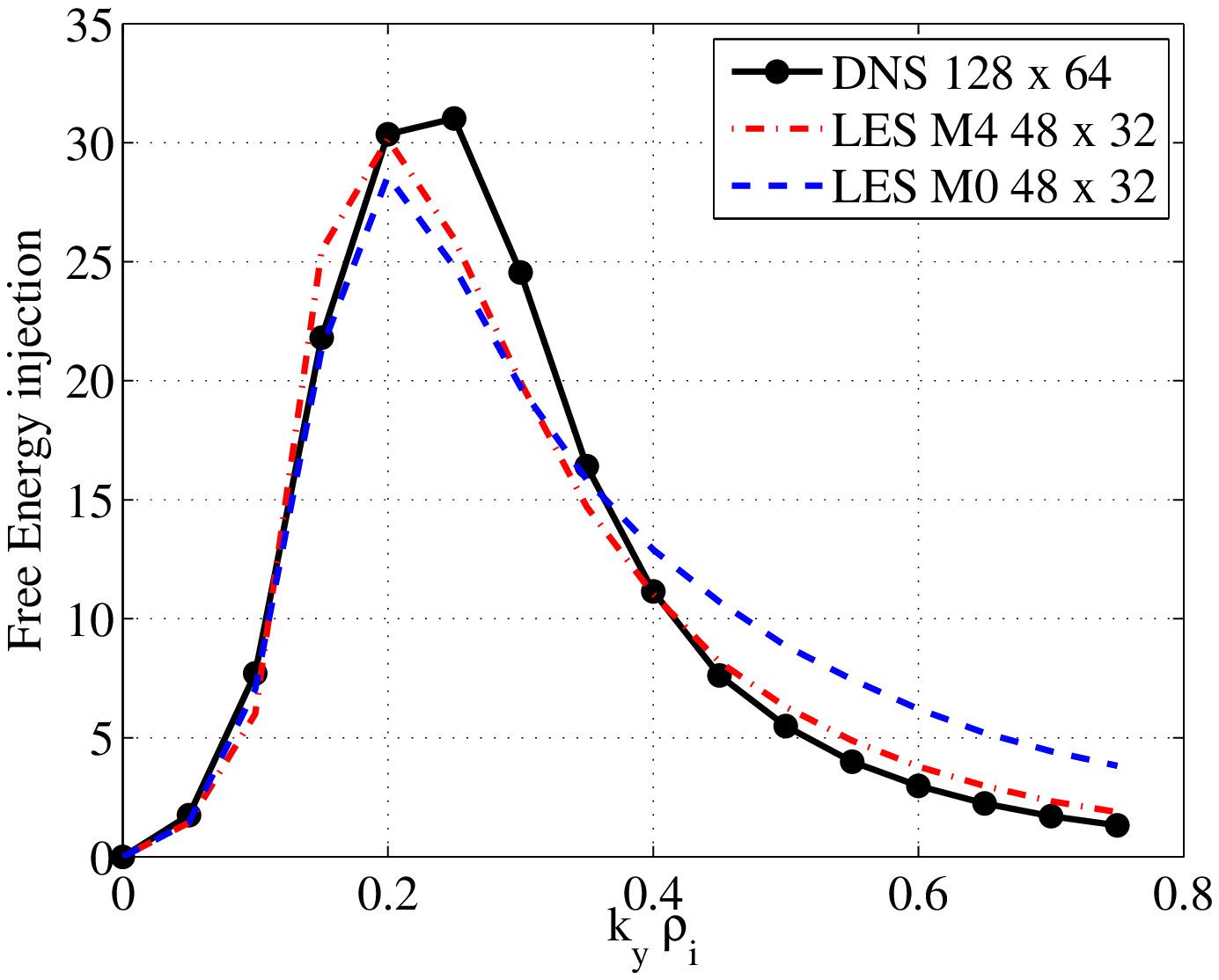}
\caption{Wavenumber spectra $\mathcal{E}^{k_y}$ (at top) and $\mathcal{G}^{k_y}$ (at bottom): Comparison between DNS and LES for the case of high shear ($\hat{s} = 1.4$).}
\label{fig:cmp_FE_ky_s14}
\end{figure}

Finally, the results for the case of high shear ($\hat{s} = 1.4$) are displayed in Fig.~\ref{fig:cmp_FE_ky_s14}. Although the turbulence level is slightly lower than for CBC parameters, the free energy and free energy injection spectra very similar to the CBC ones. A very satisfying agreement between LES and DNS is found regarding both the total free energy ($\mathcal{E}^{\textrm{\tiny M4}} = 1.23 \, \mathcal{E}^{\textrm {\tiny DNS}}$) and the total heat flux ($\mathcal{Q}^{\textrm{\tiny M4}} = 1.04 \, \mathcal{Q}^{\textrm {\tiny DNS}}$). Again, without using a model the accumulation of free energy at small scales leads to larger differences: $\mathcal{E}^{\textrm {\tiny M0}} = 1.98 \, \mathcal{E}^{\textrm {\tiny DNS}}$ and $\mathcal{Q}^{\textrm {\tiny M0}} = 1.15 \, \mathcal{Q}^{\textrm {\tiny DNS}}$.

In all three cases, an important consequence of the use of a LES model is that it prevents the accumulation of free energy at small scales. In addition, for low magnetic shear, unphysical features at small $k_y$ (which may develop due to the relative coarseness of the chosen grid) are strongly reduced. Although not shown explicitly, the $k_x$ spectra of free energy $\mathcal{E}^{k_x}$ and free energy injection $\mathcal{G}^{k_x}$ are always in good agreement. One notes that the total free energy appears to be a very sensitive diagnostic. The presence of a model considerably enhance the agreement between DNS and coarser simulations. The total heat flux is estimated with an encouraging relative error of less than $20\%$ for all cases.

\begin{figure}
\begin{center}
\includegraphics[width=8.5cm]{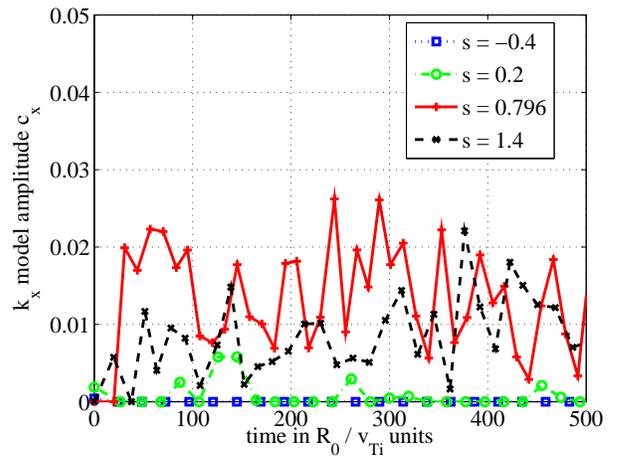}
\end{center}
\caption{Model amplitude $c_x$ as a function of time for different values of the magnetic shear $\hat{s}$.}
\label{fig:cx-cy-time-shear-scan}
\end{figure}

In contrast to the findings of the temperature gradient scan, the LES model amplitudes exhibit substantial variations for changes in the magnetic shear. In particular, the value of $c_x$ is found to be close to zero in the reversed shear case ($c_x \approx 5\cdot 10^{-5}$) as well as in the low shear case ($c_x \approx 2\cdot 10^{-3}$). Meanwhile, it departs from the CBC value ($c_x = 0.0140$) only moderately in the high shear case ($c_x = 0.0102$). The time traces of $c_x$ are shown in Fig.~\ref{fig:cx-cy-time-shear-scan}. On the other hand, the $c_y$ values do not vary much; one obtains $c_y = 0.0192$, $c_y = 0.0223$, $c_y = 0.0212$, and $c_y = 0.0185$ for the four values of magnetic shear (in increasing order). These results are a reflection of the effect of the magnetic shear on the turbulence, which includes the twisting of the perpendicular eddies along the magnetic field.

\section{Discussion \label{sec:discussion}}

In the present paper, a dynamic LES procedure has been applied to gyrokinetic turbulence as described by the GENE code. This approach provides an automatic calibration of the free parameters associated with dissipative GyroLES models. The dynamic procedure has been found to be robust in a wide parameter range of the logarithmic temperature gradient and the magnetic shear.

Comparisons between DNS and GyroLES simulations have been based on free energy and free energy injection spectra, $\mathcal{E}^{k_y}$ and $\mathcal{G}^{k_y}$. Generally, the use of a LES model has prevented the accumulation of free energy at small scales. While the differences regarding the total free energy can exceed $50\%$, simulations without a LES model are even much more inaccurate, exhibiting relative errors up to about $200\%$. Moreover, when considering the total heat fluxes, the GyroLES results are really encouraging, with relative errors below about $20\%$. In terms of computational cost, the GyroLES approach has been found to save a factor of about 20, requiring only about 250 CPU-hours per single run. This allows for the possibility to perform nonlinear gyrokinetic simulations an any modest cluster with relatively little effort.

Obviously, future GyroLES studies will have to take kinetic electrons and their contribution to the overall energetics into account. It may be expected that GyroLES will also be of great benefit in this wider context, again leading to major savings of computer resources. Thus, GyroLES is likely to enable large parameters scans of gyrokinetic turbulence which can be used, e.g., to efficiently couple turbulence and transport codes (see, e.g., Ref.~\cite{barnes-PoP-2010}).

\end{document}